\documentstyle[11pt]{article}

\begin{document}

\title{Singularities and the Finale of Black Hole
Evaporation\footnote{This essay received an honorable mention in the
2013 Essay Competition of the Gravity Research Foundation}}

\author{Li Xiang\thanks{xiang.lee@163.com}\\
 {\it Department of Physics}\\
 {\it Jimei University, 361021, Xiamen, Fujian province, P. R. China}\\
 Yi Ling\thanks{lingy@ihep.ac.cn}\\
 {\it Institute of High Energy Physics}\\
 {\it Chinese Academy of Sciences,
 100049,
 Beijing, P. R. China}\\
 You Gen Shen\thanks{ygshen@shao.ac.cn}\\
 {\it Shanghai Astronomical Observatory}\\
 {\it Chinese Academy of Sciences,
 200030, Shanghai, P. R. China}\\}

%\date{}
\maketitle

\begin{abstract}
In this essay we argue that once quantum gravitational effects
change the classical geometry of a black hole and remove the
curvature singularity, the black hole would not evaporate entirely
but approach a  remnant. In a modified Schwarzschild spacetime
characterized by a finite Kretschmann scalar, a minimal mass of the
black hole is naturally bounded by the existence of the horizon
rather than introduced by hand. A thermodynamical analysis discloses
that the temperature, heat capacity and the luminosity are vanishing
naturally when the black hole mass approaches the minimal value.
This phenomenon may be attributed to the existence of the minimal
length in quantum gravity. It can also be understood heuristically
by connecting the generalized uncertainty principle with the running
of Newton's gravitational constant.
\end{abstract}
\newpage

Black hole thermodynamics is a fascinating arena for the interplay
between gravity and quantum theory. An unsolved problem is whether a
black hole can evaporate entirely. Following from Hawking's
prediction\cite{hawk}, a black hole with mass $M$  can emit thermal
radiation of temperature $T=(8\pi M)^{-1}$, thus becoming lighter
and hotter. Its temperature and luminosity($\sim M^{-2}$) become
explosive when the black hole mass approaches zero. In order to
remove this un-physical result, one usually believes that quantum
gravitational effects should be taken into account when the black
hole becomes very small and its size is comparable to the Planck
length.

The black hole remnant is a well motivated expectation in
theoretical physics and astrophysics. It is one of candidates of
solution to the information loss puzzle\cite{preskill}. It also
improves the $\Lambda$CDM model and explains the quadrupole anomaly
of the CMB spectrum\cite{pschen1}. Indeed, there are some arguments
that quantum gravitational effects prevent a black hole from total
evaporation and lead to a
remnant\cite{pschen2}\cite{yling1}\cite{yling2}\cite{reuter}\cite{nicolini}\cite{myung}.
These effects include generalized uncertainty principle, modified
dispersion relation, rainbow gravity, running gravitational constant
and noncommutative geometry. However, there are several reasons to
think the existence of remnant still indefinite. Firstly, those
effects mentioned above have not yet been observed by experiments
and we do not know whether they can be derived from a complete
theory of quantum gravity in future. Secondly, there exists a
serious discrepancy in the expressions for the temperature of the
black hole at the final stage. In literature some authors pointed
out that the remnant's temperature would vanish naturally such as in
\cite{reuter}\cite{nicolini}\cite{myung}, while some  others argued
that the remnant is characterized by a Planck temperature,
 for instance in
Refs.\cite{pschen2}\cite{yling1}\cite{yling2}. Thirdly, an opposite
argument was made by Susskind  from the viewpoint of particle
physics, saying that there is no evident symmetry to prevent a black
hole from total evaporation\cite{susskind}.

In this essay the remnant puzzle is revisited by analyzing the
temperature of a black hole near the Planck scale, where quantum
gravitational effects are firstly considered in an indirect manner.
This is motivated by such a belief that spacetime singularities can
be removed by quantizing the gravitational fields, which reminds us
that the remnant puzzle may be clarified by connecting with the
singularity problem.

We here argue that a black hole would cease evaporation at a
specific mass scale and leave a remnant, if the curvature
singularity is removed by quantum gravitational effects.

Let us consider a static and spherically symmetric black hole which
has a metric with the form
\begin{eqnarray}\label{bh1}
ds^2=-(1+2\phi_1)dt^2+(1+2\phi_2)^{-1}dr^2+r^2d\Omega,
\end{eqnarray}
where $\phi_1=\phi_1(r), \phi_2=\phi_2(r)$. Setting $ds^2=0$ and
$dr/dt=0$, we find the horizon is located by\cite{ohanian}
\begin{eqnarray}
g^{11}=1+2\phi_{2}(\xi)=0.
\end{eqnarray}
In the neighborhood of the horizon, we have
\begin{eqnarray}\label{abexpress}
1+2\phi_{2}&=&2\kappa_2(r-\xi),\nonumber\\
1+2\phi_{1}&=&2\kappa_1(r-\xi)+a,
\end{eqnarray}
where
$\kappa_1=\phi_{1}^{\prime}(\xi),\kappa_2=\phi_{2}^{\prime}(\xi),
a=1+2\phi_{1}(\xi)$. Substituting (\ref{abexpress}) into
(\ref{bh1}),  we obtain the line element near the horizon as
\begin{eqnarray}
ds^2=-(\kappa_1\kappa_2
x^2+a)dt^2+dx^2,\nonumber
\end{eqnarray}
where the new coordinate is defined as
\begin{eqnarray}
x=\int\frac{dr}{\sqrt{2\kappa_2(r-\xi)}}=\sqrt{\frac{2(r-\xi)}{\kappa_2}}.\nonumber
\end{eqnarray}
When $a=0$,  the spacetime has an imaginary period,
$\beta=2\pi/\sqrt{\kappa_1\kappa_2}$. The black hole temperature is
given by
\begin{eqnarray}\label{temp1}
T=\frac{\sqrt{\kappa_1\kappa_2}}{2\pi}=\frac{\sqrt{\phi_1^{\prime}(\xi)\phi_2^{\prime}(\xi)}}{2\pi}.
\end{eqnarray}
Note here we do not require the function $\phi_1$ to be the same as
$\phi_2$. However, from above discussion it is clear that $a=0$, or
$\phi_1(\xi)=\phi_2(\xi)$ is necessary for the black hole
temperature to be well defined. Furthermore, as
$\phi_1(\xi)=\phi_2(\xi)$, we have a nontrivial surface gravity
given by $\kappa=\sqrt{\kappa_1\kappa_2}$. Otherwise the surface
gravity is identical to zero.

The black hole evaporation makes the horizon shrink, so $\xi$
becomes smaller and smaller. We see from  Eq.(\ref{temp1}) that the
terminal phase of black hole evolution is determined by the
behaviors of $\phi_{1}^{\prime}$ and $\phi_{2}^{\prime}$ at short
distance. Since quantum gravitational effects become significant,
the correction to the black hole temperature may be crucial. As a
physical requirement, the curvature singularity(of $r=0$)  should be
removed by quantum gravitational effects, which is
 expected to restrict the asymptotic behaviors of $\phi_1$ and $\phi_2$ in the
neighborhood of the center of the black hole.

For a spacetime given by the metric (\ref{bh1}), direct calculation
reveals that the Kretschmann scalar is expressed as
\begin{eqnarray}\label{krets1}
K&=&R^{\mu\nu\rho\lambda}R_{\mu\nu\rho\lambda}\nonumber\\
&=&\frac{16\phi_1^{2}}{r^4}+\frac{8\phi_1^{\prime
2}}{r^2}+\frac{8\phi_2^{\prime
2}}{w^2r^2}+\frac{(w^{\prime}\phi_2^{\prime}-2w\phi_2^{\prime\prime})^2}{w^4},
\end{eqnarray}
where $w=(1+\phi_1)/(1+\phi_2)$. On one hand, suppose that the black
hole singularity is removed by quantum gravitational effects, then
each term on the right hand side of Eq.(\ref{krets1}) should be
finite, therefore we obtain the asymptotic behaviors of $\phi_1$ and
$\phi_2$ to be
\begin{eqnarray}
\phi_1&\rightarrow& r^{l},~l\geq 2,\nonumber\\
\phi_2&\rightarrow& r^{s},~ s\geq 2,
\end{eqnarray}
as $r\rightarrow 0$. On the other hand, we require that classical
general relativity is valid  at macroscopic scale such that the
Schwarzschild geometry is recovered as $r\rightarrow\infty$, i.e.
$\phi_1=\phi_2 \rightarrow -M/r$. Following from Rolle's theorem,
there exists $\varsigma_1$ and $\varsigma_2$ in the interval
$(0,\infty)$ such that
\begin{eqnarray}\label{extreme}
\phi_1^{\prime}(\varsigma_1)=0, \phi_2^{\prime}(\varsigma_2)=0,
\end{eqnarray}
where $\varsigma_1$ and $\varsigma_2$ are expected to be of order
of the Planck length, on which quantum gravitational effects
become significant.

 From (\ref{temp1}) and (\ref{extreme}), we have $T\rightarrow 0$
as $\xi\rightarrow\varsigma_1$ or $\xi\rightarrow\varsigma_2$. In
other words,  when the size of such a black hole approaches a
specific value with an order of the Planck length, the black hole
will cease evaporation and leave a zero temperature remnant. It is
noted that our discussion here does not directly invoke any concrete
form of quantum gravitational effects or  the theoretical details of
quantum gravity, except the requirement that the black hole is
singularity free.

The remnant should possess a Planck order mass, which is compatible
with the quantum constraint on the minimal mass of an object
available to form a black hole\cite{beken}. However, such a lower
bound on the black hole mass can not be derived from the classical
Schwarzschild spacetime itself, except the bound is introduced by
hand. In other words, the classical geometry can not restrict the
black hole mass. We here propose a concrete example that the minimal
mass is naturally associated with the existence of black hole
horizon. For simplicity, we set $g_{00}g_{11}=-1$, and consider  a
modified Schwarzschild spacetime as follows
\begin{eqnarray}\label{element1}
ds^2&=&-(1+2\phi)dt^2+(1+2\phi)^{-1}dr^2+r^2d\Omega,
\end{eqnarray}
where $\phi=-(M/r)\exp(-\varepsilon)$, and
$\varepsilon=\varepsilon(r)$ remains to be determined. In order for
the curvature singularity to be removed, it requires $\varepsilon>0$
such that $\exp(-\varepsilon)$ plays a role of damping factor as
$r\rightarrow 0$.
 If the metric (\ref{element1}) describes a black hole, the radius of the
 horizon $\xi$ satisfies
 \begin{eqnarray}\label{horizon}
 1+2\phi(\xi)=0.
\end{eqnarray}
Considering $1+\varepsilon<\exp(\varepsilon)$, we have
\begin{eqnarray}
M=\frac{1}{2}\xi
e^{\varepsilon(\xi)}>\sqrt{\xi^2\varepsilon},\nonumber
\end{eqnarray}
for any $\xi$,  the right hand side of the inequality is expected to
be a constant, it gives rise to $\varepsilon\sim r^{-2}$.  So we
suppose the modified gravitational potential to be
\begin{eqnarray}\label{potential1}
\phi=-\frac{M}{r} e^{-\alpha/r^2},
\end{eqnarray}
where $\alpha$ is a constant of order of the square of Planck
length. As a result, Eq.(\ref{horizon}) becomes
\begin{eqnarray}\label{horizon1}
2M=\xi e^{\alpha/\xi^2},
\end{eqnarray}
and the horizon radius is given by
\begin{eqnarray}\label{horizon2}
\xi=2M\sqrt{\frac{\vartheta}{W(\vartheta)}},
~~\vartheta=-\frac{\alpha}{2M^2}.
\end{eqnarray}
Here $W(\vartheta)\geq -1$ is the principal branch of Lambert-W
function which can be expressed as
\begin{eqnarray}
W(\vartheta)=\sum_{n=1}^{\infty}\frac{(-n)^{n-1}}{n!}\vartheta^{n}.\nonumber
\end{eqnarray}
A real $W$ requires $\vartheta\geq -e^{-1}$, and it is necessary
 for the black hole since $\xi$ must be real. Thus we obtain
\begin{eqnarray}\label{masslowerbound}
M\geq\sqrt{\frac{e\alpha}{2}}.
\end{eqnarray}
It means a mass scale below which an object can not form a black
hole.

The  modified spacetime (\ref{element1}) with (\ref{potential1}) can
be attributed to an effective matter fluid which simulates quantum
gravitational effects. Following from Einstein's field equations
$G^{\mu}_{\nu}= 8\pi T^{\mu}_{\nu}$, the effective stress-energy
tensor is given by
\begin{eqnarray}\label{tensor}
T^{0}_{0}&=&T^{1}_{1}=\frac{\alpha\phi}{2\pi r^4},\nonumber\\
T^{2}_{2}&=&T^{3}_{3}=\frac{\alpha\phi}{4\pi
r^4}\left(\frac{2\alpha}{r^2}-3\right).
\end{eqnarray}
Comparing (\ref{tensor}) with a fluid with
 $T^{\mu}_{\nu}=\textrm{diag}(-\rho,\textmd{p}_1,\textmd{p}_2,\textmd{p}_3)$,
we obtain
\begin{eqnarray}
\rho+\sum_{i=1}^{3}\textmd{p}_i=\frac{\alpha\phi}{2\pi
r^4}\left(\frac{2\alpha}{r^2}-3\right).
\end{eqnarray}
The strong energy condition is violated, as $r<\sqrt{2\alpha/3}$.
  Substituting (\ref{potential1}) into
(\ref{krets1}) with $\phi_1=\phi_2=\phi$, the Kretschmann scalar of
the corrected spacetime is given by
\begin{eqnarray}\label{kretsch2}
K&=&\frac{16\phi^2}{r^4}+\frac{16\phi^{\prime
2}}{r^2}+4\phi^{\prime\prime
2}\nonumber\\
&=&\frac{16M^2}{r^6}e^{-2\alpha/r^2}\left(3-\frac{14\alpha}{r^2}+\frac{33\alpha^2}{r^4}-\frac{20\alpha^3}{r^6}+\frac{4\alpha^4}{r^8}\right).
\end{eqnarray}
Remarkably,  the damping factor results in $\lim_{r\rightarrow
0}K=0$, which is in contrast to the classical case  with an
unavoidable  singularity.

Next let us consider the thermodynamics  of such a modified black
hole. From Eq.(\ref{potential1}), the black hole temperature is read
as
\begin{eqnarray}
 T=\frac{\phi^{\prime}(\xi)}{2\pi}=\frac{1}{4\pi\xi}\left(1-\frac{2\alpha}{\xi^2}\right),
 \end{eqnarray}
where Eq.(\ref{horizon1}) has been considered. Accordingly, the heat
capacity and the luminosity are respectively given by
 \begin{eqnarray}
 C&=&\frac{dM}{dT}=\frac{dM/d\xi}{dT/d\xi}=-2\pi \xi^{2} \frac{(\xi^2-2\alpha)}{\xi^2-6\alpha}e^{\alpha/\xi^2},\nonumber\\
 L&=& \sigma T^4
 A=\frac{\sigma}{64\pi^3\xi^2}\left(1-\frac{2\alpha}{\xi^2}\right)^4,
 \end{eqnarray}
where $\sigma$ is the Stefan-Boltzmann constant, and $A$ is the area
of black hole horizon. Furthermore, with the use of
Eq.(\ref{horizon2}), $T, C, L$ can be rewritten as
\begin{eqnarray}
T&=&\frac{1}{8\pi M}\left(\frac{W}{\vartheta}\right)^{1/2}(1+W),\nonumber\\
C&=&-8\pi M^2\left(\frac{\vartheta}{W}\right)^{1/2}\frac{1+W}{1+3W},\nonumber\\
L&=&\frac{\sigma}{256\pi^3M^2}\left(\frac{W}{\vartheta}\right)(1+W)^4.
\end{eqnarray}
 These expressions return to the well known results  for Schwarzschild black hole
  as $\vartheta=0$, with the use of $W(0)=0$ and $W^{\prime}(0)=1$. Besides the minimal mass given
 by (\ref{masslowerbound}), there is  another special value for the mass in
 the modified spacetime, i.e.
 \begin{eqnarray}
 M_c=\sqrt{\frac{3\alpha}{2e^{1/3}}},
 \end{eqnarray}
 below which  $W<-1/3$ and $C>0$. This means a process of
 quantum cooling evaporation\cite{myung}. When the mass approaches the minimal value($\vartheta\rightarrow -1/e$), $W\rightarrow -1$,  and then $T\rightarrow 0,~
C\rightarrow
 0,~ L\rightarrow 0$, which means a stable remnant.

 Next we  attempt to explore the quantum effects of gravity which would potentially lead to the modification of the metric with
a potential as in Eq.(\ref{potential1}). In the context of quantum
gravity, it is widely believed that the standard uncertainty
principle in quantum theory
 should be generalized to be
 compatible with the belief that there exists a minimal length
 to be observed in the lab, which is also viewed as the fundamental nature of quantum
 spacetime\cite{gup}\cite{kempf}\cite{ahlu}. Such a generalized uncertainty principle
would lead to a modification of the Heisenberg's commutation
relation
 as
 \begin{eqnarray}\label{gup1}
 [p,x]=-iz.
 \end{eqnarray}
where $z=z(p)$. For distinction, $[p,q]=-i$ denotes the  usual
Heisenberg's commutation relation.  Eq.(\ref{gup1}) leads to the
uncertainty relation $\Delta x\geq\langle z\rangle/{\Delta p}$,
where the  right hand side of the inequality is required to be not
less than Planck order length. For instance, in \cite{hh} Harbach
and Hossenfelder suggest
\begin{eqnarray}\label{hhgup}
z=e^{\alpha^{\prime} p^2},
\end{eqnarray}
which means a minimal length of order of $\sqrt{\alpha^{\prime}}$
and provides a natural  UV regulator in quantum field theory. On the
other hand,  observing Eq.(\ref{horizon2}), we notice that the black
hole remnant possesses a horizon of the size $\xi=\sqrt{2\alpha}$,
which restricts the minimal region that the outer observers can
probe. This implies that the presence of the potential in
(\ref{potential1}) may be interpreted as an effect due to the
generalized uncertainty principle(GUP). If so, then our mechanism
presented here is helpful to get out of trouble from
Ref.\cite{pschen2}, therein GUP gives rise to a black hole remnant
of Planck temperature. Next we will look into this point by
investigating the dynamics of a particle moving in a spacetime with
quantum fluctuations.

It is easy to understand that the dynamics of a particle moving in
the classical spacetime should be different from that in a
background dominated by quantum fluctuations. Consequently, GUP as
an important effects of quantum spacetime should be responsible for
the modification of the particle dynamics. The relation between GUP
and the corrected spacetime may be established by considering  the
impact of the GUP on the motion of particles  in an indirect and
heuristic manner as we will describe as follows. Once the modified
dynamics is known, one can infer the metric form of the corrected
spacetime with a reasonable sense.

Let us consider a test particle moving in a weak gravitational field
described by Newton potential
\begin{eqnarray}
\phi_{N}=-\frac{G_N M}{r},\nonumber
\end{eqnarray}
where $G_N$ is the Newton gravitational constant. The  motion of the
particle obeys Newton's second law, and the acceleration can be
expressed as
\begin{eqnarray}\label{operator1}
a_N=[p,q]^{-1}[p,\phi_N],
\end{eqnarray}
in the context of quantum mechanics. When Heisenberg commutation
relation is generalized to (\ref{gup1}), Eq.(\ref{operator1}) is
changed to
\begin{eqnarray}\label{operator2}
a_N\rightarrow a&=&[p,x]^{-1}[p,\phi_N]\nonumber\\
&=&iz^{-1}[p,\phi_N]\nonumber\\
&=&[p,q]^{-1}[p,\phi_{p}],
\end{eqnarray}
where $\phi_p=z^{-1}\phi_N$.  Comparing (\ref{operator1}) with
(\ref{operator2}), we deduce that the influence of GUP can be
described by an effective potential as follows
\begin{eqnarray}
\phi_{p}=-\frac{G M}{r},
\end{eqnarray}
where the effective Newton gravitational constant is defined as
\begin{eqnarray}\label{rg}
 G=G(z)=z^{-1}G_N,
\end{eqnarray}
 which is running.  An argument for Eq.(\ref{rg}) can also be made in a Gedanken experiment that the Colella-Overhause-Werner effect is reexamined by the
GUP deduced wave-particle duality\cite{lix}, where the GUP's
influence on the gravity-induced phase shift can be attributed to
the running Newton constant, and meanwhile Heisenberg principle is
retained. Setting $G_N=1$ and considering Eq.(\ref{hhgup}), we
obtain $G=z^{-1}=e^{-\alpha^{\prime} p^2}$. As an observable, $p^2$
should take the average value, $p^2\rightarrow\langle p^2\rangle$,
and $\langle p^2\rangle\geq(\Delta p)^2\geq 1/(\Delta q)^2$, where
$\Delta q$ denotes the position uncertainty of the particle. It is
necessary to require $\Delta q\leq r$, otherwise one can not
distinct the test particle from the field source, and  then the
position of the particle will lose its meaning. Consequently, we
obtain
\begin{eqnarray}\label{potential2}
\phi_p=-\frac{M}{r} e^{-\alpha^{\prime}p^2}\geq-\frac{M}{r}
e^{-\alpha^{\prime}/r^2}.
\end{eqnarray}
The right hand side of inequality describes a quasi-classical
gravitational field, since it is associated with the minimal
uncertainty($r\Delta p\sim 1$) and has an analogy to the coherent
states in quantum mechanics. Furthermore,  it has the same form as
the corrected potential (\ref{potential1}). This is consistent with
our expectation: the corrected spacetime represented by
(\ref{potential1}) is related to the existence of the minimal
length, and  can be understood by connecting the GUP with the
running $G$, at least in the weak field approximation. The effective
potential (\ref{potential2}) may suffer a modification in a strong
gravitational field. However, the qualitative conclusion should not
be changed, if $G$ remains a decreasing function of $p$ and the
damping factor can be expressed as $\exp(-1/r^{s}), s>0$.

It is surprising and interesting that GUP is associated with the
running $G$ via Eq.(\ref{rg}), since they are based on different
considerations. For instance,  a running $G$ can be derived from the
renormalization group equation in an effective theory of quantum
gravity\cite{reuter}. On the other hand, following from
(\ref{hhgup}) and (\ref{rg}), if $\alpha^{\prime} p^2<<1$, the
running Newton constant reads as $G\simeq G_N(1+\alpha^{\prime}
p^2)^{-1}$, which is similar to the result presented in
\cite{reuter}. This similarity  implies a subtle relation between
GUP and the renormalization group method,  a fundamental framework
associated with various aspects of quantum gravity.

In the end of this essay we point out that our mechanism presented
here should also be applicable to the early universe. It is
expectable that the initial cosmological singularity would be
removed in an indirect manner. A relevant interesting issue is how
the GUP influences the inflationary cosmology. Eq.(\ref{rg})
provides a guide by which we can consider various generalizations of
Heisenberg principle and their applications to gravitation physics.
Following from the deductions of the running $G$,  we may judge the
reliability of  those generalized uncertainty relations appeared in
literature.


\begin{thebibliography}{99}
\bibitem{hawk}S. W. Hawking,  Commun. Math. Phys. 43(1975), 199.
\bibitem{preskill} Do black hole destroy information,
hep-th/9209058.
\bibitem{pschen1} F. Scardigli, C. Gruber, and P. Chen, Phys. Rev.
D83(2011), 0635707.
\bibitem{pschen2}R. J. Adler, P. Chen and  D. I. Santiago,  Gen. Rel. Grav. 33 (2001), 2101.
\bibitem{yling1} Y. Ling, B. Hu and X. Li, Phys. Rev. D73(2006),
087702.
\bibitem{yling2} Y. Ling, X. Li, and H. B. Zhang, Mod. Phys. Letts.
A22(2007), 2749.
\bibitem{reuter} A. Bonanno and M. Reuter, Phys. Rev.
D62(2000),043008.
\bibitem{nicolini} P. Nicolini, A. Smailagic and E. Spallucci, Phys.
Letts. B632(2006), 547.
\bibitem{myung} Y. S. Myung, Y-W. Kim and Y-J. Kim, Phys. Letts.
B656(2007), 221.
\bibitem{susskind} Susskind's viewpoint was mentioned in
Ref.\cite{pschen2}.
\bibitem{ohanian} H. C. Ohanian and R. Ruffini, Gravitaion and Spacetime(2nd ed.),
W.W. Norton and Company, Inc. 1994.
\bibitem{beken}J. D. Benkenstein, Phys. Rev. D7(1973), 2333.
\bibitem{gup}L. J. Garay, Int. J. Mod. Phys. A10(1995), 145.
\bibitem{kempf}A. Kempf, G. Mangano and R.B. Mann,   Phys.
Rev.  D52(1995), 1108.
\bibitem{ahlu}D.V. Ahluwalia, Phys. Letts. A275(2000), 31.
\bibitem{hh}U. Harbach and S. Hossenfelder, Phys.Lett. B632(2006),
379.
\bibitem{lix} Li Xiang and L. F. Xu, in preparation.
\end{thebibliography}
\end{document}